%% file: Main.tex
\documentclass[sigconf]{acmart}

\renewcommand\footnotetextcopyrightpermission[1]{} 
\AtBeginDocument{
 \providecommand\BibTeX{{%
 \normalfont B\kern-0.5em{\scshape i\kern-0.25em b}\kern-0.8em\TeX}}}
\usepackage{soul}
\usepackage{algorithm} 
\usepackage{algpseudocode}
\usepackage{algorithmicx}
\usepackage{float}

\usepackage{pifont}

\usepackage{todonotes} 

\usepackage{xcolor}

\input{math_commands.tex}

\usepackage{amsmath}

\begin{document}

\title{DRAM-Profiler: An Experimental DRAM RowHammer Vulnerability Profiling Mechanism \vspace{-0.4em}}

\author{Ranyang Zhou$^\dagger$, Jacqueline T. Liu$^\ddagger$, Nakul Kochar$^{\dagger}$, Sabbir Ahmed$^{\ddagger}$, \\Adnan Siraj Rakin$^\ddagger$, and Shaahin Angizi$^\dagger$}
\affiliation{
\institution{\small$^\dagger$Department of Electrical and Computer Engineering, New Jersey Institute of Technology, Newark, NJ, USA}
\institution{\small$^\ddagger$Department of Computer Science, State University of New York at Binghamton, NY \country{USA}}}
\email{arakin@binghamton.edu, shaahin.angizi@njit.edu} \vspace{-1em}

\begin{abstract}
RowHammer stands out as a prominent example, potentially the pioneering one, showcasing how a failure mechanism at the circuit level can give rise to a significant and pervasive security vulnerability within systems. Prior research has approached RowHammer attacks within a static threat model framework. Nonetheless, it warrants consideration within a more nuanced and dynamic model. This paper presents a low-overhead DRAM RowHammer vulnerability profiling technique termed DRAM-Profiler, which utilizes innovative test vectors for categorizing memory cells into distinct security levels. The proposed test vectors intentionally weaken the spatial correlation between the aggressors and victim rows before an attack for evaluation, thus aiding designers in mitigating RowHammer vulnerabilities in the mapping phase. While there has been no previous research showcasing the impact of such profiling to our knowledge, our study methodically assesses 128 commercial DDR4 DRAM products. The results uncover the significant variability among chips from different manufacturers in the type and quantity of RowHammer attacks that can be exploited by adversaries. \vspace{-0.5em}
\end{abstract} 


\maketitle
\pagestyle{plain}

\vspace{-1em}
\section{Introduction}
\begin{figure}[t]
\begin{center}
\begin{tabular}{c}
\includegraphics [width=0.94\linewidth]{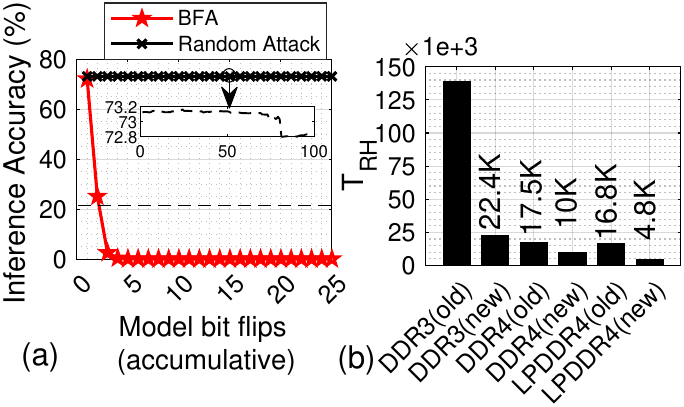}\vspace{-0.4em}
 \end{tabular} \vspace{-0.8em}
\caption{(a) Targeted bit-flipping vs. random bit-flipping for an 8-bit quantized ResNet-34 on ImageNet~\cite{deng2009imagenet} dataset, (a) RowHammer thresholds \cite{woo2022scalable,marazzi2023rega}.}\vspace{-1.1em}
\label{RHthre}
\end{center}
\end{figure}

Recent research has demonstrated that adversaries can exploit the RowHammer vulnerability present in DRAM to systematically and precisely manipulate bits across diverse applications, including proficiently trained neural networks, resulting in a notable impact on accuracy \cite{hong2019terminal,rakin2019bit}. Illustrated in Fig. \ref{RHthre}(a), such so-called Bit-Flip Attacks (BFAs) can reduce the accuracy of an 8-bit quantized ResNet-34 on the ImageNet dataset from 73.1\% to 0\% by targeting only 5 bits. Fig. \ref{RHthre}(b) reports that the RowHammer threshold has experienced a notable decline in recent years. For instance, on LPDDR4 (new), the attacker requires approximately 4.5 times fewer Hammer Counts (HC)
compared to DDR3 (new) \cite{woo2022scalable}. This threshold is anticipated to nearly vanish with the introduction of DDR5 \cite{marazzi2023rega}.

To effectively mitigate RowHammer attacks, comprehensive investigation, and analysis of pertinent influencing factors are imperative. As research progresses, Error Correction Code (ECC) techniques \cite{mutlu2019rowhammer,lee2019twice} have been developed across various directions to combat RowHammer attacks. Nonetheless, we cannot get detailed parameters of the organization from manufacturers because they defend the secrecy of chip structures, which means even the same RowHammer attack model has different performance to different chips. System manufacturers such as Apple \cite{Apple} and HP \cite{HP} commonly employ a standard RowHammer mitigation approach by elevating the refresh rate, albeit at the cost of significant power consumption and susceptibility to compromise \cite{mutlu2019rowhammer}. Intel's pTRR \cite{kaczmarski2014thoughts} and various research work propose a proactive strategy involving the monitoring of row activations, termed HC. The memory controller tracks HC and initiates refresh cycles on victim rows once the number of row activations surpasses a predefined Maximum Activate Count (MAC) threshold ($T_{MAC}$), typically stored on the Serial Presence Detect (SPD) chip within the DRAM module \cite{frigo2020trrespass}.

Previous studies have predominantly addressed RowHammer attacks under a static threat model, emphasizing fixed parameters. However, we advocate for a more sophisticated and adaptable approach, acknowledging the evolving nature of security threats. In contrast to static models, a dynamic framework accommodates the fluidity of attack vectors and defense mechanisms, thus providing a more comprehensive understanding of RowHammer vulnerabilities. By embracing this nuanced perspective, researchers can better anticipate emerging threats and devise effective countermeasures to safeguard against RowHammer and similar exploits. In this work, we introduce DRAM-Profiler, a novel technique for profiling DRAM RowHammer vulnerabilities with minimal overhead. DRAM-Profiler employs innovative test vectors to classify memory cells into different security levels. The main contributions of this paper are as follows:

(1) We demonstrate that the bit-flip induced by RowHammer attacks is intricate and variable, necessitating varied analyses associated with different patterns applied in the RowHammer attack model; (2) We propose a comprehensive classification of DRAM cells referred to as \textit{cell's security level} within the chip to enhance the visibility of the impact of RowHammer attacks; and (3) Our experimental findings reveal substantial variability in the robustness of cells across 128 chips sourced from 7 major DRAM manufacturers. Consequently, we recommend the adoption of targeted defense mechanism designs as a more effective approach.

\section{Overview}
\begin{figure}[t]
\begin{center}\vspace{-0.5em}
\begin{tabular}{c}
\includegraphics [width=0.97\linewidth]{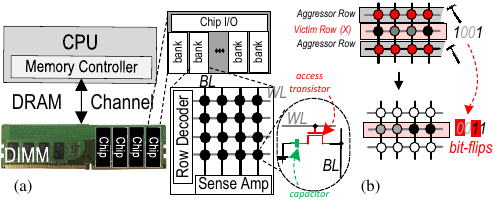}\vspace{-0.4em}
 \end{tabular} \vspace{-0.8em}
\caption{(a) Hierarchical organization of DRAM, (b) Double-sided RowHammer attack \cite{kim2014flipping}.}\vspace{-1em}
\label{DRAM} \vspace{-0.2em}
\end{center}
\end{figure}

\textbf{DRAM Organization \& Commands.} The DRAM chip is a hierarchical structure consisting of several memory banks, as shown in Fig. \ref{DRAM}(a). Each bank comprises 2D sub-arrays of memory bit-cells virtually ordered in memory matrices, with billions of DRAM cells on modern chips. Each DRAM bit-cell consists of a capacitor and an access transistor. The charge status of the bit-cell's capacitor is used to represent binary ``1'' or ``0'' \cite{zhou2022red}. In idle mode, the memory controller turns off all enabled DRAM rows by sending the Precharge (\texttt{PRE}) command on the command bus. This will precharge the Bit-Line (BL) voltage to $\frac{V_{DD}}{2}$. In the active mode,
the memory controller will send an Activate (\texttt{ACT)} command to
the DRAM 
module to activate the Word-Line (WL). Then, all DRAM cells connected to the WL share their charges with the corresponding BL. Through this process, BL voltage deviates from the precharged $\frac{V_{DD}}{2}$. The sense amplifier then senses this deviation and amplifies it to $V_{DD}$ or 0 in the row buffer. The memory controller can then send read (\texttt{RD})/write (\texttt{WR}) commands to transfer data from/to the sense amplifier array.

\noindent\textbf{DRAM Timing Parameters.}
In the context of DRAM standards, a comprehensive array of timing parameters is established, with each parameter prescribing the minimum temporal separation between two successive DRAM commands to uphold seamless operational integrity. The most basic parameter is the clock cycle ($t_{CK}$) used to measure all parameters. Row Active Time (\textbf{$t_{RAS}$}) encompasses the temporal window demarcating an \texttt{ACT} command and the subsequent \texttt{PRE} command. During this prescribed $t_{RAS}$ interval, the restoration of charge within the DRAM cells on the open DRAM row is effectuated to ensure optimal performance. Row Precharge Time (\textbf{$t_{RP}$}) signifies the temporal gap between the issuance of a \texttt{PRE} command and the subsequent \texttt{ACT} command. The imposition of $t_{RP}$ is instrumental in closing the open WL and initiating the pre-charging of the DRAM BLs to the voltage level of $\frac{V_{DD}}{2}$. Retention time in DRAM refers to the duration for which a memory cell can hold its stored data without requiring a refresh operation. It can be influenced by various factors, including the density of cells, electromagnetic interference, and so on. The critical timing parameters are fundamental in ensuring the reliable and efficient operation of DRAM modules across different standards. In the RowHammer model, the retention time of certain victim rows may experience a substantial reduction. The Refresh Window ($t_{REFW}$) is essentially the interval within which all DRAM cells must be refreshed to prevent data loss or corruption.

\noindent\textbf{RowHammer in DDR4 \& Protection Mechanisms.}
Kim et al. \cite{kim2014flipping} were the pioneers in conducting an extensive study on the characteristics of RowHammer bit-flips in DDR3 modules. They observed that approximately 85\% of the tested modules were susceptible to RowHammer attack. Therefore, the majority of earlier RowHammer research is centered on DDR3 systems \cite{seaborn2015exploiting}. With the prospect of having a RowHammer-less landscape, DDR4 modules have been introduced. While there are documented instances of RowHammer on DDR4 chips in previous studies \cite{lipp2020nethammer,gruss2018another}, these findings pertain to earlier generations of DDR4.  To the best of our knowledge, the only recent and established work exploring the multi-sided fault injection model is TRRespass \cite{frigo2020trrespass}.

Multiple software and hardware mitigation mechanisms have been proposed to reduce the impact of RowHammer-based attacks \cite{kim2014flipping,marazzi2022protrr,zhou2023dnn,zhou2022lt,zhou2023dram,zhou2023p}. 
The hardware-based research efforts can be classified into two categories, i.e., \textit{victim-focused} mechanism with probabilistic refreshing (e.g., PRA \cite{kim2014architectural}, PARA \cite{kim2014flipping}, ProHIT \cite{son2017making}, ProTRR \cite{marazzi2022protrr}) and \textit{aggressor-focused} mechanism by counting activations (e.g.,  TRR \cite{hassan2021uncovering}, Hydra \cite{qureshi2022hydra}, CBT \cite{seyedzadeh2018mitigating}, Panopticon \cite{bennett2021panopticon}, CRA \cite{kim2014architectural}, TWiCe \cite{lee2019twice}, Graphene \cite{park2020graphene}, Mithril \cite{kim2022mithril}).
The system manufacturers tend to follow the mechanisms that explicitly detect RowHammer conditions and intervene, such as increasing refresh rates and access counter-based approaches. 
Along this line, Target Row Refresh (TRR) \cite{frigo2020trrespass} and counter-based detection methods \cite{kim2014architectural,qureshi2022hydra,seyedzadeh2016counter} require add-on hardware to calculate rows' activation and record it to other fast-read-memory (SRAM \cite{lee2019twice}/CAM \cite{park2020graphene}). The controller will then refresh the target row if the number reaches MAC \cite{frigo2020trrespass}.
TWiCe \cite{lee2019twice} is a per-row counter-based RowHammer prevention solution based on the idea that the number of ACTs within $t_{REFW}$ is limited. Instead of detecting the rows, TWiCe only checks the number of \texttt{ACT}s. However, inserting a counter for each memory row imposes a substantial burden both from latency and power consumption perspectives \cite{seyedzadeh2016counter}. To tackle this issue, recent works \cite{seyedzadeh2016counter} consider the storage of counters in a dedicated section of DRAM or use a set-associative counter cache implemented within the memory controller to enhance the efficiency of accessing frequently utilized counters \cite{kim2014architectural}.
CAT \cite{seyedzadeh2018mitigating} is a counter-based solution that keeps track of the number of \texttt{ACT}s performed on a set of DRAM rows and initiates a refresh operation for the entire group of rows, once the HC reaches the MAC. The counter-based solutions have been enabled by adding a new DRAM command called Nearby Row Refresh (NRR) \cite{lee2019twice,park2020graphene}
that will be issued to refresh the relevant victim rows.

The JEDEC standard outlines three potential configurations for the MAC value: (1) unlimited, if the DRAM module claims to be RowHammer-free; (2) untested, if the DRAM module has not undergone post-production inspection; or (3) $T_{MAC}$ indicating the specific number of \texttt{ACT}s the DRAM module can withstand (e.g., 1M). It has been revealed \cite{frigo2020trrespass} that, irrespective of the DRAM manufacturer, the majority of DDR4 modules assert unlimited MAC value. \vspace{-0.5em}

\section{DRAM-Profiler}
\subsection{Formulating the Problem}
A bit-flip occurs exclusively when there is a disparity in the bit values of adjacent rows. This raises the query regarding the differentiation of data among DRAM rows, a consequence of manufacturers' topology techniques. Consequently, the likelihood of adjacent rows differing from the target row on every bit is exceedingly low, resulting in numerous bits within the victim row sharing identical values with those in the adjacent row. According to this hypothesis, certain bits remain immune to flipping when adversaries employ an SG attack strategy.
Nonetheless, in the DB attack model, the scenario becomes intricate. Ideally, the two assailant rows would exhibit diversity, each contrasting with the victim row on every bit. However, in specific instances, the sheer abundance of distinct bits complicates this ideal scenario.

\begin{figure}[t]
\begin{center}
\begin{tabular}{c}
\includegraphics [width=0.84\linewidth]{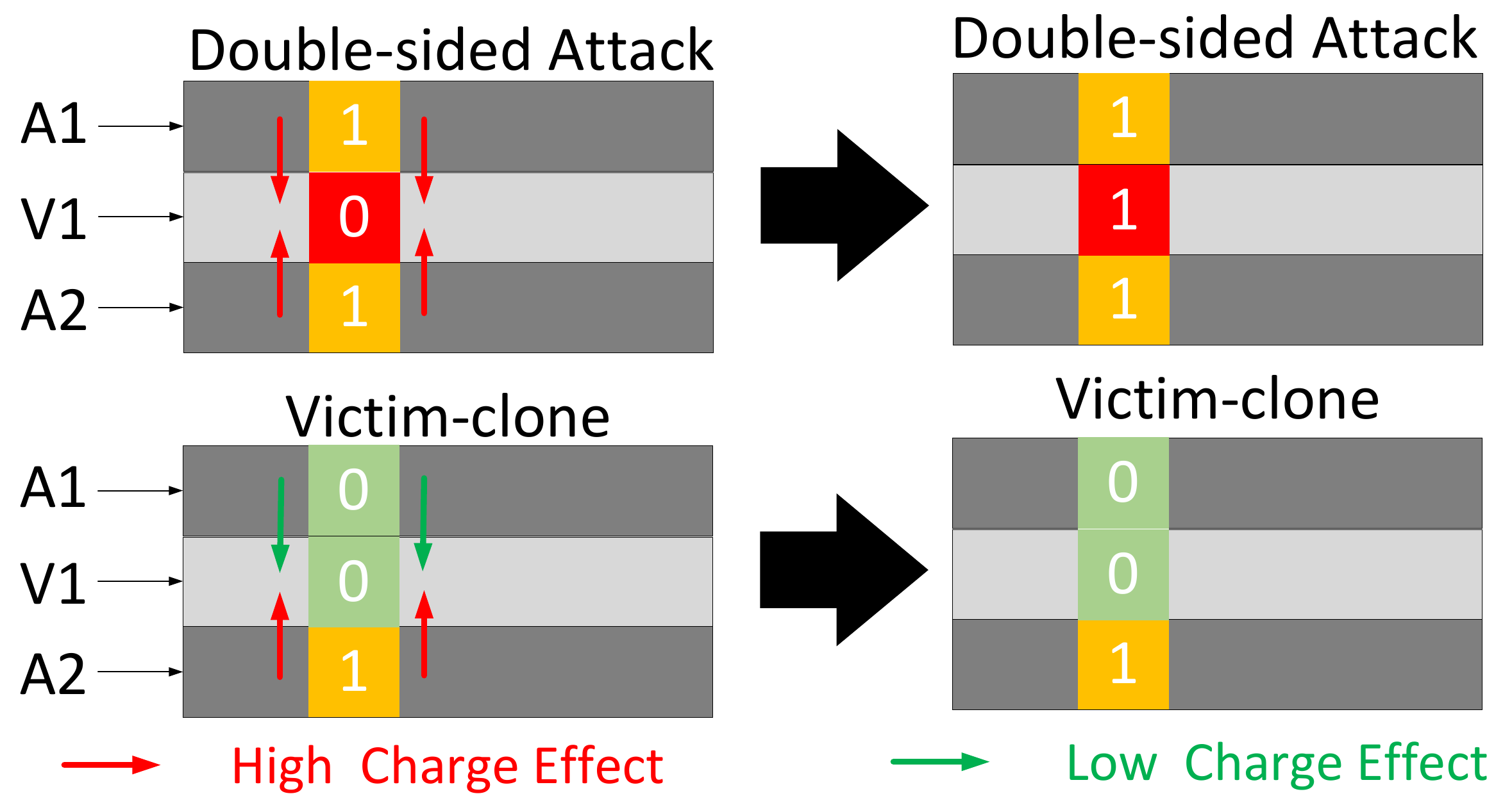}\vspace{-0.4em}
 \end{tabular} \vspace{-0.8em}
\caption{Single-Sided (SG), Double-Sided (DB), and Victim-Clone (VC) attack models.}
\label{MI} \vspace{-1.8em}
\end{center}
\end{figure}

Previous studies \cite{lee2019twice,jattke2022blacksmith,seyedzadeh2016counter} have overlooked comparable specifics, and their assessment of RowHammer relies on analyzing the subsequent conditions: $(i)$ complete dissimilarity between all bits of the attack row and the victim row, and $(ii)$ conducting experiments using real DRAM storage data. However, owing to technical disparities among various manufacturers, this data pattern can be perceived as random. To enhance comprehension of the factors contributing to bit-flipping in RowHammer attacks, we decided to create a new research model
As shown in Fig. \ref{MI}, we assume the Double-sided (DB) RowHammer attack is based on the ideal case where each bit of the aggressor rows (A1 \& A2) differs from the victim (V1). Based on our speculation and analysis of previous research works \cite{kim2020revisiting}, we hypothesize that both cells in the attacking row exert a significant charging effect on the cells in the victim row. According to the previous research in \cite{mutlu2019rowhammer}, we know that the RowHammer vulnerability relies on rapid and repeated access to adjacent rows, causing electromagnetic interference that can lead to bit-flips in the victim row. However, for bit-flips to occur, there must be a difference in the charge state between the aggressor and victim rows.  

Victim-Clone (VC) is our proposed attack model to make the victim row suffer less when the DRAM is under the DB attack. Leveraging this model, we can focus on a more detailed study of the effects of leakage between cells and prolong the stability of cells within the victim's row, preventing them from experiencing bit-flips for an extended period. The VC model essentially copies the victim row to one of the aggressor rows to ensure that each bit of the victim row is only affected by one adjacent flipped bit. As shown in Fig. \ref{MI}, in this model, the cell in A1 has a low charge effect, and that in A2 has a high charge effect. 

\vspace{-0.5em}
\subsection{DRAM Security Level}
\begin{figure}[t]
\begin{center}
\begin{tabular}{c}
\includegraphics [width=0.95\linewidth]{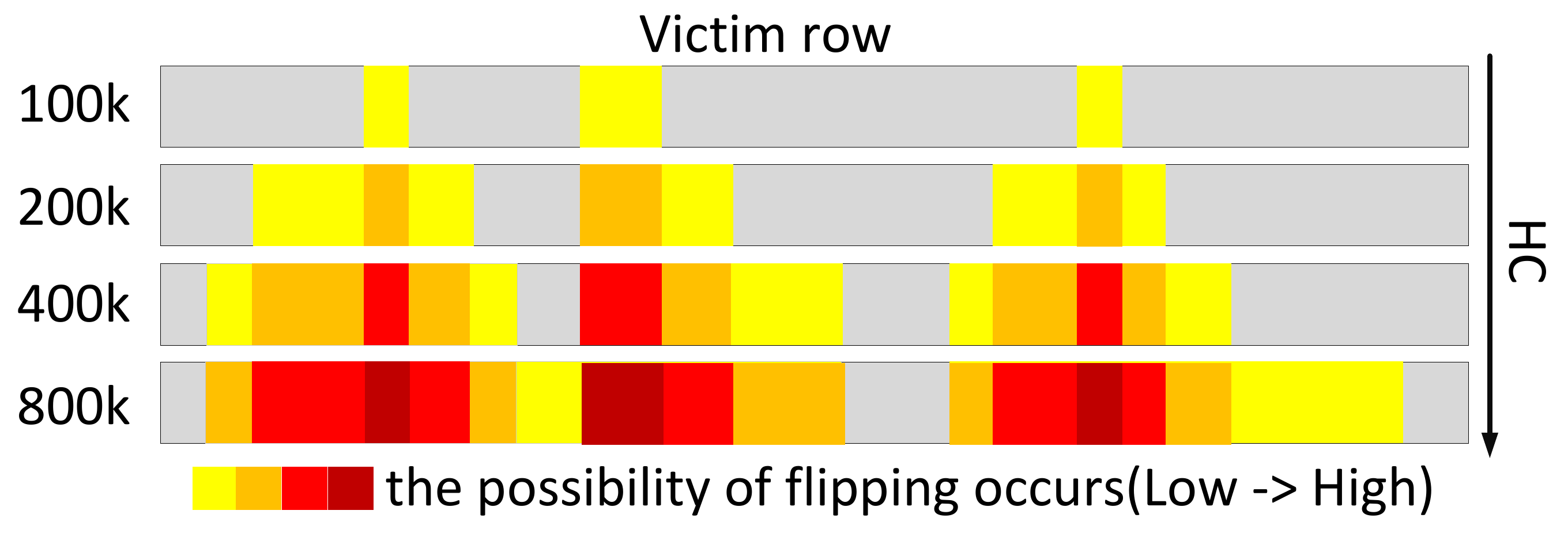}\vspace{-0.4em}
 \end{tabular} \vspace{-0.8em}
\caption{The vulnerability of  cells associated with the HC}\vspace{-0.4em}
\label{CL} \vspace{-1.1em}
\end{center}
\end{figure}

Based on the findings reported in our preceding study [Omitted for review purposes], a direct correlation exists between the increment of HC 
and the observed rise in bit-flip occurrences in DRAM
This phenomenon signifies an escalating number of cells susceptible to charge leakage as HC values increase. Alternatively, a granular examination of individual HC values unveils distinct patterns in cell presence across different levels. Some cells demonstrate consistent presence across multiple HC levels, while others exhibit sporadic or negligible presence. Utilizing a color-coded scheme to represent DRAM cell frequencies at varying HC levels allows for the visualization of these patterns, facilitating a more comprehensive understanding of DRAM vulnerability to RowHammer attacks. As shown in Fig. \ref{CL}, assume we collect samples from the same chip subjected to RowHammer attacks at various HC
levels. In every tier, we emphasize the cells where bit-flips occur. As the HC escalates from upper to lower tiers, the number of these cells generating bit-flips rises. Concurrently, we note that cells causing bit-flips at lower HC levels persist in generating bit-flips at higher HC levels, implying their consistency across varying HC levels. Therefore, the more frequently a flipped cell appears at all levels, the more vulnerable it is. In other words, if some cells appear in different HC levels simultaneously, the highlighted color will be darker. Thus, the color bars in Fig. \ref{CL} can represent the vulnerability of cells. The four colors from left to right (from bright yellow to dark red) represent the degree of vulnerability from low to high.
This model can be exploited for the following reasons:

$(i)$ To empirically analyze significant variability among chips from different manufacturers. Consequently, we aim to investigate whether this discrepancy correlates with the quantity of highly vulnerable cells within the chip.

$(ii)$  To investigate variations in the rate at which the number of bit-flips increases with rising HC levels. Therefore, this model facilitates a more detailed examination of the differences between cells from different manufacturers.

$(iii)$ To explore RowHammer
attack modes yield outcomes. This model enables us to evaluate the resilience of cells from different manufacturers to various attack modes.
\vspace{-0.5em}

\section{Experiments}\vspace{-0.3em}
In this section, we will design and conduct experiments to validate the hypotheses posited in the preceding section.\vspace{-0.3em}
\subsection{Prerequisites}
\begin{figure}[t]
\begin{center}
\begin{tabular}{c}
\includegraphics [width=0.7\linewidth]{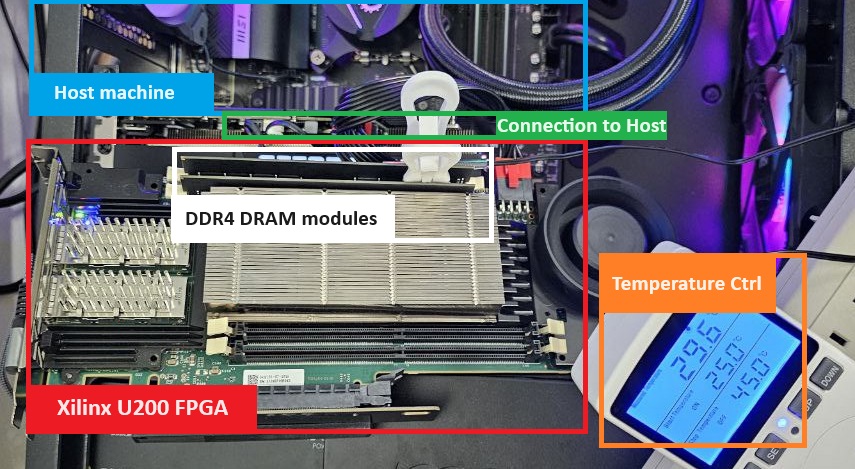}\vspace{-0.7em}
 \end{tabular} \vspace{-0.4em}
\caption{Our testing infrastructure for DDR4 modules.}
\label{frame}
\vspace{-1em}
\end{center}
\end{figure}

\textbf{Framework Setup \& Testing Infrastructure.} We test the DRAM chips by extensively modifying the DRAM-Bender \cite{olgun2023dram} to have a versatile FPGA-based DRAM attack exploration framework for DDR4 with an in-DRAM compiler API installed on our host machine. Our testing infrastructure, as shown in Fig. \ref{frame}, consists of the Alveo U200 Data Center Accelerator Card \cite{Alevo} as the FPGA that accepts DDR4 modules and runs the test programs based on Algorithm \ref{alg} by sending DDR4 command traces generated by the host machine. The key idea is to take control of memory modules for DDR4 interfaces with straightforward high-level programming to test, characterize, and run the generated programs on the host machine. The driver is designed to send instructions across the PCIe bus to the FPGA to be stored on the board.
Besides, to have a fair comparison among various under-test DRAM chips, the temperature is kept below 30$^{\circ}$C with INKBIRDPLUS 1800W temperature controller.

\noindent\textbf{Minimizing Interference.} Before implementing the proposed attack scenario, DRAM refresh \cite{JEDEC} and rank-level ECC are disabled to minimize their interference with RowHammer bit-flips. However, proprietary RowHammer protection techniques (e.g., Target Row Refresh \cite{frigo2020trrespass,hassan2021uncovering}) are in place.

\noindent\textbf{Chips Tested.} 
To profile DRAM cell vulnerabilities, the experiments are conducted on a range of 128 commercialized DRAM chips from eight different manufacturers (mf.) as listed in Table \ref{DRAMchips} with various die densities and die revisions. 
\vspace{-1.2em}

\begin{table}[h]
  \centering
  \caption{Under-test DRAM chips.} \vspace{-1.2em}
\scalebox{0.75}{
\begin{tabular}{lcccll}
\hline
\multicolumn{1}{c}{\textbf{Vendor}} & \textbf{\#Chips} & \textbf{Freq (MHz)}  & \textbf{Die rev.}    & \textbf{Org.} & \textbf{Date} \\ \hline
mf-A (Crucial 16 GB)                 & 16               & 3200                 &   C                   &x8              & N/A             \\
mf-B (Kingston 16GB)               & 16               & 2666    & G              &x8 & 2152            \\ 
mf-C (Micron 16GB)                  & 16               & 2133                 &  B                    &x4               &  2126              \\ 
mf-D (NEMIX 16GB)                 & 16               & 2133                 &B                      &x4               &   1733              \\
mf-E (SK Hynix 16GB)                & 16               & 2400                 &A                      &x8               & 1817              \\
mf-F (Patriot Viper 16GB)                 & 16               & 3600                 & C                    &x8               &  N/A             \\ 
mf-G (Samsung 16GB)                 & 16               & 2400                 & B                     & x8                 &  2053          \\

\hline
\end{tabular}}
  \label{DRAMchips} 
\end{table}
\vspace{-2em}
\subsection{Programming}
We devise the following RowHammer test procedure in Algorithm 1 to conduct experiments, aimed at characterizing various DDR4 DRAM modules and implementing three fault injection models by enabling control over the HC. We start by initializing both the FPGA and the testing framework to introduce disruptions in DRAM timing, enabling the insertion of instructions into the DRAM chip. Then, we initialize the row address denoted by $Initial\_Row$, whereby the data pattern will be allocated in line-4, and load the prepared $Data\_pattern$ and $Data\_pattern\_inv$ in line-5. We set all bits in $Data\_pattern$ and $Data\_pattern\_inv$ to ``1'' 
and ``0'', respectively. This initialization establishes a straightforward state, facilitating a clearer observation of the attack intensity. The initialization is completed at this point. We will further allocate different patterns according to different cases. In the case of Single-sided attack (SG) and DB, we utilize the traditional allocation method to guarantee that the aggressor row and the victim row store data in a precisely opposite manner. In the case of the VC, one of the aggressor rows replicates the data from the victim row, while the other row contains data that is the opposite, effectively simulating the duplication of the victim row onto one of the aggressor rows. Lines 9, 14, and 19 provide the hammering instructions applied to the DRAM rows. We hammer $Initial\_Row+1$ in SG and $Initial\_Row$ \& $Initial\_Row+2$ in both VC and DB models. Following the execution of all RowHammer reaching the pre-set HC, we retrieve the data from all rows. Ultimately, we ascertain the number of bit-flips by comparing the read data with the initially stored data.\vspace{-1em}
\begin{algorithm}[h]
        \caption{\small RowHammer test procedure for SG, VC, and DB}
          \scalebox{0.65}{
    \begin{minipage}{3\linewidth}
        \begin{algorithmic}[1]
            \State $\textbf{Procedure: \textit{RHtest}}$
                \State $\textbf{Input} \hspace{4pt} HC$ 
                \State $\textbf{Initialize} \hspace{4pt} FPGA()\hspace{4pt}\&\hspace{4pt}Platform()$ 
                \State $\textbf{Define} \hspace{4pt} Inital\_Row$ 
                \State $Load \hspace{4pt} Data\_pattern \hspace{4pt} $\&$ \hspace{4pt} Data\_pattern\_inv $
                \State $\textbf{Case SG:}$ 
                \State $\hspace{16pt} Initial\_Row \gets Data\_pattern$
                \State $\hspace{16pt} Initial\_Row+1 \gets Data\_pattern\_inv$
                \State $\hspace{16pt} Hammer \hspace{4pt} Initial\_Row+1 \hspace{4pt} for(HC)$
                \State $\textbf{Case VC:}$
                \State $\hspace{16pt} Initial\_Row \gets Data\_pattern$
                \State $\hspace{16pt} Initial\_Row+1 \gets Data\_pattern\_inv$
                \State $\hspace{16pt} Initial\_Row+2 \gets Data\_pattern\_inv$
                \State $\hspace{16pt} Hammer \hspace{4pt} Initial\_Row \hspace{4pt} \& \hspace{4pt} Initial\_Row+2 \hspace{4pt} for(HC)$
                \State $\textbf{Case DB:}$
                \State $\hspace{16pt} Initial\_Row \gets Data\_pattern$
                \State $\hspace{16pt} Initial\_Row+1 \gets Data\_pattern\_inv$
                \State $\hspace{16pt} Initial\_Row+2 \gets Data\_pattern$
                \State $\hspace{16pt} Hammer \hspace{4pt} Initial\_Row \hspace{4pt} \& \hspace{4pt} Initial\_Row+2 \hspace{4pt} for(HC)$

        \State $Receive\_Data(Platform);\hspace{16pt}//Write\hspace{4pt}data\hspace{4pt}back\hspace{4pt}to\hspace{4pt}hostPC$
        \State $Detect\_BitFlips(Victim\_Row)$

    \State \textbf{end} $\textbf{Procedure}$
        \end{algorithmic} 
         \end{minipage} \vspace{-3em}}
         \label{alg}
    \end{algorithm}
\vspace{-2em}
\subsection{Analysis of the Results}
\begin{figure*}[t]
\begin{center}
\begin{tabular}{c}
\includegraphics [width=0.99\linewidth]{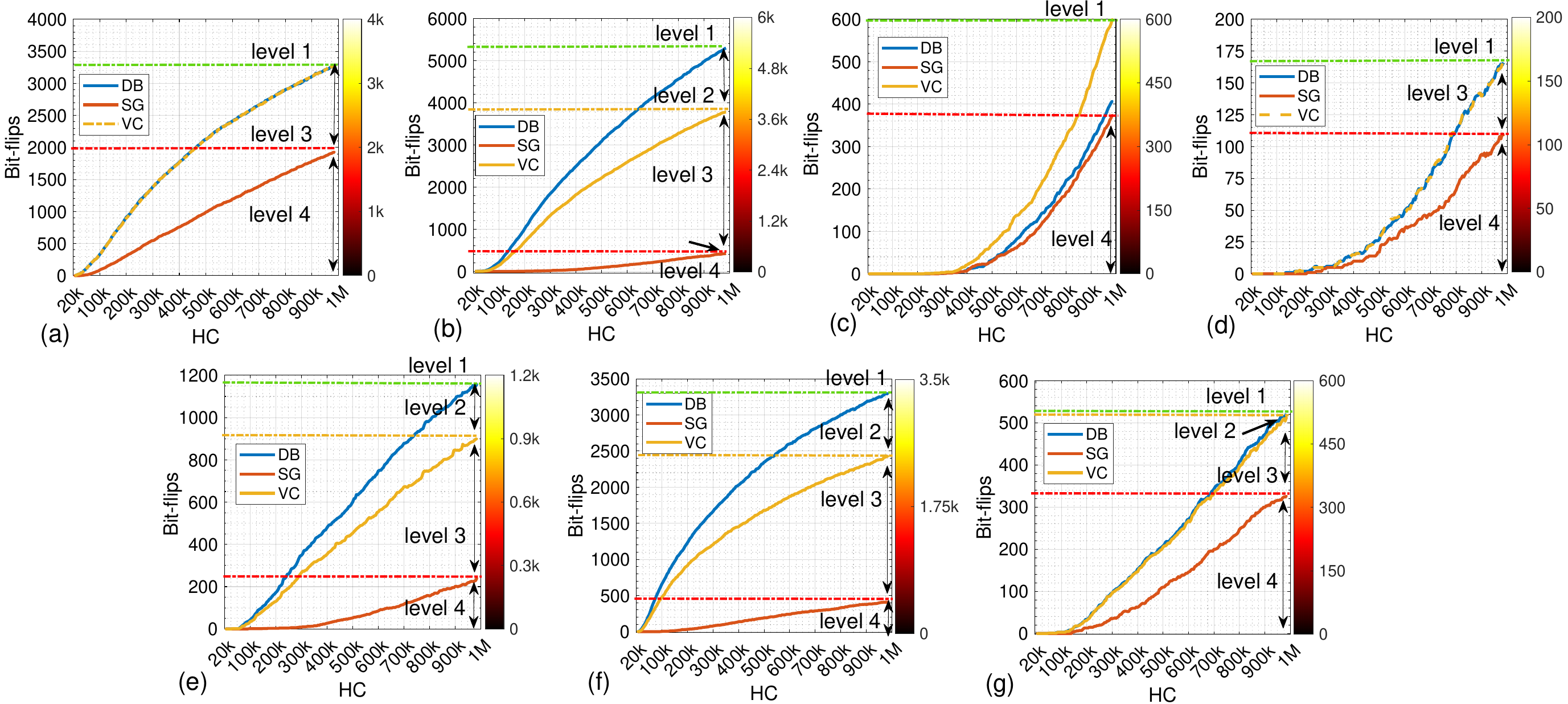}\vspace{-1.4em}
 \end{tabular}
\caption{Analysis of the security levels of cells on (a) mf-A, (b) mf-B, (c) mf-C, (d) mf-D, (e) mf-E, (f) mf-F, (g) mf-G.}
\label{result} \vspace{-0.5em}
\end{center}
\end{figure*}

Fig. \ref{result} represents the comprehensive analysis results of the security levels of DRAM cells. In every plot, there are three curves for different RowHammer attack models, i.e., DB,
SG, and VC. The X-axis denotes HC, and the Y-axis represents the number of cells at which bit-flip occurred. The typical $t_{RAS}$ values for DDR4 memory modules can range from approximately 36 to 48 $t_{CK}$ \cite{choi2020reducing}, although these values may vary depending on the module's speed rating (e.g., DDR4-2133, DDR4-2400, DDR4-3200, etc.). For example, the duration of a clock cycle for DDR4-2400 memory can be calculated as $t_{CK}=\frac{1}{2400MT/s}$. In our design, each $t_{RAS}$ comprises three components: \texttt{ACT}, \texttt{Sleep}, and \texttt{PRE}, where \texttt{Sleep} is set to 5$t_{CK}$. In order to more accurately emulate real-world scenarios, we set a maximum limit of 1M for the HC.
Given that we suspended the DRAM refresh command in this experiment, it became necessary to manually account for retention time. So in a refresh window ($t_{REF}$) the maximum number of HC must be less than $\frac{t_{REF}}{t_{RAS}}$=1.37M. Practically, the application cannot be composed entirely of activations, so we limit the number of activations used for RowHammer to 1M. Here we list our key observations regarding the under-test chips.

\vspace{0.5em}
\hspace{-0.5em}\fcolorbox{black}{gray!50}{\begin{minipage}{24em}
\textbf{Obs.\#1.} Compared with DB model, VC model cannot effectively reduce the number of bit-flips. 
\end{minipage}}\vspace{0.5em}

As discussed, the VC attack model is a way to make the victim row less vulnerable by copying the victim row to one of the aggressor rows. Therefore, the number of bit-flips produced by the final cell should be hypothetically close to that created by the SG
attack. However, based on the empirical findings, it is evident that only chips from three manufacturers exhibit improvement following the implementation of the VC. As depicted in Fig. \ref{result}(b)(e)(f), upon reaching the HC limit (1M), the bit-flips induced by VC decreased by approximately 25\% compared to DB yet remained over four times more than those induced by SG. This observation contradicts our initial hypothesis: within the DB model, replicating the victim row onto one of the aggressor rows does not significantly decrease the frequency of bit-flips.
We can draw a new conclusion from this: when the attacker ensures that one bit in the aggressor rows differs from the victim row, they can efficiently flip the one in the victim row. Confirming the prior reports, the DB attack is more likely to produce a bit-flip than the SG attack. As shown in Fig. \ref{result}(a)(d)(g), the results of VG and DB almost overlap, meaning that these cells can produce bit-flip as long as at least one cell in the adjacent row differs from it. Fig. \ref{result}(c) represents a special case, in which the VC model generates more bit-flips than DB. The factors contributing to this observation remain unknown.

\vspace{0.5em}
\hspace{-0.5em}\fcolorbox{black}{gray!50}{\begin{minipage}{24em}
\textbf{Obs.\#2.}  Various cells demonstrate diverse levels of resistance to various attack models. 
\end{minipage}}\vspace{0.5em}

Undoubtedly, within the same chip, certain cells are susceptible to RowHammer attacks, whereas others remain unaffected. However, determining the susceptibility of a cell poses a challenge. To address this, we employ a visual approach for classification. We have opted to use a four-level scale, ranging from level 1 to level 4, to denote the extent of cell vulnerability. Lower levels indicate a lower likelihood of bit-flips, while higher levels suggest greater susceptibility. Take Fig. \ref{result}(b)(e)(f) as examples, among these chips, we posit that if a cell succumbs to SG, it can be deemed the most vulnerable to attack. Consequently, when HC is 1M, we classify all cells that induce bit-flips as level 4. Next, we consider cells that do not induce bit-flips in SG but exhibit them in VC. We categorize these cells as level 3. As previously discussed, if cells with high vulnerability manifest bit-flips in low-threat attacks, they are also likely to experience bit-flips in high-threat attacks. Consequently, when HC is 1M, we derive the level 3 count by subtracting the total number of bit-flips in VC from the total number of bit-flips in SG. Applying the same principle, we classify cells exhibiting behaviors between DB and VC as level 2. Finally, if cells withstand even the DB attack, we classify them as level 1. Excluding the chips from these three manufacturers, as shown in Fig. \ref{result}(a)(d)(g), due to the scarcity of cells between DB and VC, we delete level 2 and keep level 3. Figure \ref{result}(c) presents an exception; we cannot classify level 2 and level 3 using the previous rules. Therefore, in this scenario, we can only classify cells as level 1 and level 4.

\vspace{0.5em}
\hspace{-0.5em}\fcolorbox{black}{gray!50}{\begin{minipage}{24em}
\textbf{Obs.\#3.} Tailored DRAM protection mechanisms, designed according to specific chip topologies, will be necessary and more efficient.
\end{minipage}}\vspace{0.5em}

From our experiments, we discovered significant variations in RowHammer attacks across chips from different manufacturers, likely due to distinct manufacturing processes. Consequently, we contend that designing tailored defense mechanisms based on the specific characteristics of individual chips may yield greater effectiveness. For example, considering Fig. \ref{result}(c)(d)(g), which has a significant proportion of level 1 cells, it may opt to employ a defense strategy targeting levels 3, 2, and 1. In Fig. \ref{result}(b)(e)(f), the primary characteristic is the exceedingly low number of cells in level 4, coupled with a larger number of cells in levels 3 and 2. As such, implementing a defense mechanism against DB attacks could be appropriate. Finally, in Fig. \ref{result}(a), all the levels are average, so the counter-based defense mechanisms can be recommended. While there are multiple approaches to defending against RowHammer attacks, the most straightforward method involves identifying the factors that render cells deferentially vulnerable to attack. Enhancing these influencing factors will represent the most effective defense against RowHammer attacks.

\vspace{0.5em}
\hspace{-0.5em}\fcolorbox{black}{gray!50}{\begin{minipage}{24em}
\textbf{Obs.\#4.}  The stability of cells in chips varies among different manufacturers.
\end{minipage}}\vspace{0.5em}

Here, we introduced a cell classification method, yet the stability of cells also influences our classification to some extent. Stability refers to the fluctuation range in the number of cells that induce bit-flips when HC is at a specific value. A broader range indicates lower stability. For instance, in real-world scenarios, cells may occasionally trigger bit-flips once HC reaches a particular value due to interference from various factors. However, in experimental settings, no bit-flips occur. Fig. \ref{result}(d) and (g) serve as prime examples of low stability. We observe that the curves for these two chips exhibit irregular fluctuations, indicating significant variability in the number of bit-flips at certain HC values. In comparison, other chips are relatively stable.
\vspace{-1em}

\subsection{DNN Weight Attack}
To further analyze the effectiveness of the conducted study in DNN application, we incorporate the three different attack models/levels, i.e., SG, VC, and DB, into the popular BFA attack framework~\cite{rakin2019bit,yao2020deephammer} via only targeting cells that will succumb to the corresponding attack levels, and conduct the adjusted BFA attack on a quantized ResNet-20~\cite{he2015delving} trained on CIFAR-10~\cite{krizhevsky2014cifar}. Table \ref{DNN atttack} displays the number of iterations needed to degrade model accuracy to a random guess level (i.e., 10\%) under the three distinct attack strategies across all under-test DRAM chips.
We observe that the numbers of required iterations vary extraordinarily across different chips. Echoing the observations from Fig. \ref{result}(a)(d)(g), where the curves of VC and DB overlap, the number of required iterations are identical (15, 27, and 19, respectively). Generally, a single-sided row hammer requires more bit-flips to achieve the attacker's objective on most chips. \vspace{-1em}

\begin{table}[h]
\centering
\caption{Number of required iterations for BFA attack~\cite{rakin2019bit} to degrade a quantized ResNet-20~\cite{he2015delving} trained on CIFAR-10~\cite{krizhevsky2014cifar} to a random guess level.} \vspace{-1em}
\scalebox{0.65}{
\begin{tabular}{lcccll}
\hline
\multicolumn{1}{c}{\textbf{Vendor}} & \textbf{Single-sided attack}  & \textbf{Victim-Clone attack} & \textbf{Double-sided attack} \\ \hline
mf-A &18  &15  &15    \\
mf-B &50  &17  &14    \\ 
mf-C &20  &46  &55    \\ 
mf-D &40  &27  &27    \\
mf-E &74  &34  &20    \\
mf-F &49  &14  &15    \\ 
mf-G &28  &19  &19    \\
\hline
\end{tabular}}
\label{DNN atttack} \vspace{-1em}
\end{table}

\vspace{-1em}
\section{Conclusion}
This paper introduces a mechanism, DRAM-Profiler, for experimental DRAM RowHammer vulnerability profiling. This mechanism is proposed to make the analysis of the RowHammer attack model more comprehensive and visible. We explore various RowHammer models to reintroduce a more authentic setting, addressing a previously overlooked aspect in prior research. The revised model provides a more nuanced understanding of performance variations across different manufacturers' chips, highlighting the necessity for a dynamic, rather than static, approach to the RowHammer problem. Additionally, our investigation delved into the phenomenon of cell flipping triggered by varying activation frequencies. This led us to classify cells based on their activation frequency, as the crux of bit-flip occurrences lies in charge alterations within the cells. By categorizing cells in this manner, we can effectively gauge the resilience of different chips against specific activation frequencies. Armed with these insights, we are poised to develop more targeted defense strategies against RowHammer attacks in future designs.

\bibliographystyle{ACM-Reference-Format}
\bibliography{sample-base}

\end{document}

%% file: math_commands.tex
\usepackage{amsmath,amsfonts,bm}

\def\eqref#1{equation~\ref{#1}}

\def\1{\bm{1}}

\DeclareMathAlphabet{\mathsfit}{\encodingdefault}{\sfdefault}{m}{sl}
\SetMathAlphabet{\mathsfit}{bold}{\encodingdefault}{\sfdefault}{bx}{n}














%% file: Main.bbl

\begin{thebibliography}{39}


\ifx \showCODEN    \undefined \def \showCODEN     #1{\unskip}     \fi
\ifx \showDOI      \undefined \def \showDOI       #1{#1}\fi
\ifx \showISBNx    \undefined \def \showISBNx     #1{\unskip}     \fi
\ifx \showISBNxiii \undefined \def \showISBNxiii  #1{\unskip}     \fi
\ifx \showISSN     \undefined \def \showISSN      #1{\unskip}     \fi
\ifx \showLCCN     \undefined \def \showLCCN      #1{\unskip}     \fi
\ifx \shownote     \undefined \def \shownote      #1{#1}          \fi
\ifx \showarticletitle \undefined \def \showarticletitle #1{#1}   \fi
\ifx \showURL      \undefined \def \showURL       {\relax}        \fi
\providecommand\bibfield[2]{#2}
\providecommand\bibinfo[2]{#2}
\providecommand\natexlab[1]{#1}
\providecommand\showeprint[2][]{arXiv:#2}

\bibitem[App(2015)]%
        {Apple}
 \bibinfo{year}{2015}\natexlab{}.
\newblock \bibinfo{booktitle}{\emph{Apple, Inc. About the security content of mac efi security update 2015-001.}}
\newblock
\urldef\tempurl%
\url{https://support.apple.com/en-au/HT204934.}
\showURL{%
\tempurl}


\bibitem[HP(2015)]%
        {HP}
 \bibinfo{year}{2015}\natexlab{}.
\newblock \bibinfo{booktitle}{\emph{HP, Inc. Hp moonshot component pack.}}
\newblock
\urldef\tempurl%
\url{https://support.hpe.com/hpsc/doc/public/ display?docId=c04676483,May2015.}
\showURL{%
\tempurl}


\bibitem[JED(2020)]%
        {JEDEC}
 \bibinfo{year}{2020}\natexlab{}.
\newblock \bibinfo{booktitle}{\emph{JESD79-4C: DDR4 SDRAM Standard}}.
\newblock
\urldef\tempurl%
\url{https://www.xilinx.com/products/boards-and-kits/alveo.html}
\showURL{%
\tempurl}


\bibitem[Ale(2021)]%
        {Alevo}
 \bibinfo{year}{2021}\natexlab{}.
\newblock \bibinfo{booktitle}{\emph{Xilinx Inc., Xilinx Alveo U200 FPGA Board}}.
\newblock
\urldef\tempurl%
\url{https://www.xilinx.com/products/boards-and-kits/alveo.html}
\showURL{%
\tempurl}


\bibitem[Bennett et~al\mbox{.}(2021)]%
        {bennett2021panopticon}
\bibfield{author}{\bibinfo{person}{Tanj Bennett} {et~al\mbox{.}}} \bibinfo{year}{2021}\natexlab{}.
\newblock \showarticletitle{Panopticon: A complete in-dram rowhammer mitigation}. In \bibinfo{booktitle}{\emph{DRAMSec}}, Vol.~\bibinfo{volume}{22}. \bibinfo{pages}{110}.
\newblock


\bibitem[Choi et~al\mbox{.}(2020)]%
        {choi2020reducing}
\bibfield{author}{\bibinfo{person}{Haerang Choi} {et~al\mbox{.}}} \bibinfo{year}{2020}\natexlab{}.
\newblock \showarticletitle{Reducing DRAM refresh power consumption by runtime profiling of retention time and dual-row activation}.
\newblock \bibinfo{journal}{\emph{MICPRO}}  \bibinfo{volume}{72} (\bibinfo{year}{2020}).
\newblock


\bibitem[Deng et~al\mbox{.}(2009)]%
        {deng2009imagenet}
\bibfield{author}{\bibinfo{person}{Jia Deng} {et~al\mbox{.}}} \bibinfo{year}{2009}\natexlab{}.
\newblock \showarticletitle{Imagenet: A large-scale hierarchical image database}. In \bibinfo{booktitle}{\emph{2009 IEEE conference on computer vision and pattern recognition}}. Ieee, \bibinfo{pages}{248--255}.
\newblock


\bibitem[Frigo et~al\mbox{.}(2020)]%
        {frigo2020trrespass}
\bibfield{author}{\bibinfo{person}{Pietro Frigo} {et~al\mbox{.}}} \bibinfo{year}{2020}\natexlab{}.
\newblock \showarticletitle{TRRespass: Exploiting the many sides of target row refresh}. In \bibinfo{booktitle}{\emph{SP}}. IEEE, \bibinfo{pages}{747--762}.
\newblock


\bibitem[Gruss et~al\mbox{.}(2018)]%
        {gruss2018another}
\bibfield{author}{\bibinfo{person}{Daniel Gruss} {et~al\mbox{.}}} \bibinfo{year}{2018}\natexlab{}.
\newblock \showarticletitle{Another flip in the wall of rowhammer defenses}. In \bibinfo{booktitle}{\emph{SP}}. IEEE, \bibinfo{pages}{245--261}.
\newblock


\bibitem[Hassan et~al\mbox{.}(2021)]%
        {hassan2021uncovering}
\bibfield{author}{\bibinfo{person}{Hasan Hassan} {et~al\mbox{.}}} \bibinfo{year}{2021}\natexlab{}.
\newblock \showarticletitle{Uncovering in-dram rowhammer protection mechanisms: A new methodology, custom rowhammer patterns, and implications}. In \bibinfo{booktitle}{\emph{MICRO}}. \bibinfo{pages}{1198--1213}.
\newblock


\bibitem[He et~al\mbox{.}(2015)]%
        {he2015delving}
\bibfield{author}{\bibinfo{person}{Kaiming He}, \bibinfo{person}{Xiangyu Zhang}, \bibinfo{person}{Shaoqing Ren}, {and} \bibinfo{person}{Jian Sun}.} \bibinfo{year}{2015}\natexlab{}.
\newblock \showarticletitle{Delving deep into rectifiers: Surpassing human-level performance on imagenet classification}. In \bibinfo{booktitle}{\emph{ICCV}}. \bibinfo{pages}{1026--1034}.
\newblock


\bibitem[Hong et~al\mbox{.}(2019)]%
        {hong2019terminal}
\bibfield{author}{\bibinfo{person}{Sanghyun Hong} {et~al\mbox{.}}} \bibinfo{year}{2019}\natexlab{}.
\newblock \showarticletitle{Terminal Brain Damage: Exposing the Graceless Degradation in Deep Neural Networks Under Hardware Fault Attacks.}. In \bibinfo{booktitle}{\emph{USENIX}}. \bibinfo{pages}{497--514}.
\newblock


\bibitem[Jattke et~al\mbox{.}(2022)]%
        {jattke2022blacksmith}
\bibfield{author}{\bibinfo{person}{Patrick Jattke} {et~al\mbox{.}}} \bibinfo{year}{2022}\natexlab{}.
\newblock \showarticletitle{Blacksmith: Scalable rowhammering in the frequency domain}. In \bibinfo{booktitle}{\emph{SP}}. IEEE, \bibinfo{pages}{716--734}.
\newblock


\bibitem[Kaczmarski(2014)]%
        {kaczmarski2014thoughts}
\bibfield{author}{\bibinfo{person}{Marcin Kaczmarski}.} \bibinfo{year}{2014}\natexlab{}.
\newblock \bibinfo{title}{Thoughts on intel xeon e5-2600 v2 product family performance optimisation--component selection guidelines}.
\newblock
\newblock


\bibitem[Kim et~al\mbox{.}(2014a)]%
        {kim2014architectural}
\bibfield{author}{\bibinfo{person}{Dae-Hyun Kim} {et~al\mbox{.}}} \bibinfo{year}{2014}\natexlab{a}.
\newblock \showarticletitle{Architectural support for mitigating row hammering in DRAM memories}.
\newblock \bibinfo{journal}{\emph{IEEE CAL}} \bibinfo{volume}{14}, \bibinfo{number}{1} (\bibinfo{year}{2014}), \bibinfo{pages}{9--12}.
\newblock


\bibitem[Kim et~al\mbox{.}(2020)]%
        {kim2020revisiting}
\bibfield{author}{\bibinfo{person}{Jeremie~S Kim} {et~al\mbox{.}}} \bibinfo{year}{2020}\natexlab{}.
\newblock \showarticletitle{Revisiting rowhammer: An experimental analysis of modern dram devices and mitigation techniques}. In \bibinfo{booktitle}{\emph{ISCA}}. IEEE, \bibinfo{pages}{638--651}.
\newblock


\bibitem[Kim et~al\mbox{.}(2022)]%
        {kim2022mithril}
\bibfield{author}{\bibinfo{person}{Michael~Jaemin Kim} {et~al\mbox{.}}} \bibinfo{year}{2022}\natexlab{}.
\newblock \showarticletitle{Mithril: Cooperative row hammer protection on commodity dram leveraging managed refresh}. In \bibinfo{booktitle}{\emph{HPCA}}. IEEE, \bibinfo{pages}{1156--1169}.
\newblock


\bibitem[Kim et~al\mbox{.}(2014b)]%
        {kim2014flipping}
\bibfield{author}{\bibinfo{person}{Yoongu Kim} {et~al\mbox{.}}} \bibinfo{year}{2014}\natexlab{b}.
\newblock \showarticletitle{Flipping bits in memory without accessing them: An experimental study of DRAM disturbance errors}.
\newblock \bibinfo{journal}{\emph{ACM SIGARCH Computer Architecture News}} \bibinfo{volume}{42}, \bibinfo{number}{3} (\bibinfo{year}{2014}), \bibinfo{pages}{361--372}.
\newblock


\bibitem[Krizhevsky et~al\mbox{.}(2014)]%
        {krizhevsky2014cifar}
\bibfield{author}{\bibinfo{person}{Alex Krizhevsky}, \bibinfo{person}{Vinod Nair}, {and} \bibinfo{person}{Geoffrey Hinton}.} \bibinfo{year}{2014}\natexlab{}.
\newblock \showarticletitle{The CIFAR-10 dataset}.
\newblock \bibinfo{journal}{\emph{online: http://www. cs. toronto. edu/kriz/cifar. html}}  \bibinfo{volume}{55} (\bibinfo{year}{2014}).
\newblock


\bibitem[Lee et~al\mbox{.}(2019)]%
        {lee2019twice}
\bibfield{author}{\bibinfo{person}{Eojin Lee} {et~al\mbox{.}}} \bibinfo{year}{2019}\natexlab{}.
\newblock \showarticletitle{TWiCe: Preventing row-hammering by exploiting time window counters}. In \bibinfo{booktitle}{\emph{ISCA}}. \bibinfo{pages}{385--396}.
\newblock


\bibitem[Lipp et~al\mbox{.}(2020)]%
        {lipp2020nethammer}
\bibfield{author}{\bibinfo{person}{Moritz Lipp} {et~al\mbox{.}}} \bibinfo{year}{2020}\natexlab{}.
\newblock \showarticletitle{Nethammer: Inducing rowhammer faults through network requests}. In \bibinfo{booktitle}{\emph{EuroS\&PW}}. IEEE, \bibinfo{pages}{710--719}.
\newblock


\bibitem[Marazzi et~al\mbox{.}(2022)]%
        {marazzi2022protrr}
\bibfield{author}{\bibinfo{person}{Michele Marazzi} {et~al\mbox{.}}} \bibinfo{year}{2022}\natexlab{}.
\newblock \showarticletitle{Protrr: Principled yet optimal in-dram target row refresh}. In \bibinfo{booktitle}{\emph{SP}}. IEEE, \bibinfo{pages}{735--753}.
\newblock


\bibitem[Marazzi et~al\mbox{.}(2023)]%
        {marazzi2023rega}
\bibfield{author}{\bibinfo{person}{Michele Marazzi} {et~al\mbox{.}}} \bibinfo{year}{2023}\natexlab{}.
\newblock \showarticletitle{REGA: Scalable Rowhammer Mitigation with Refresh-Generating Activations}. In \bibinfo{booktitle}{\emph{SP}}. IEEE.
\newblock


\bibitem[Mutlu and Kim(2019)]%
        {mutlu2019rowhammer}
\bibfield{author}{\bibinfo{person}{Onur Mutlu} {and} \bibinfo{person}{Jeremie~S Kim}.} \bibinfo{year}{2019}\natexlab{}.
\newblock \showarticletitle{Rowhammer: A retrospective}.
\newblock \bibinfo{journal}{\emph{IEEE TCAD}}  \bibinfo{volume}{39} (\bibinfo{year}{2019}).
\newblock


\bibitem[Olgun et~al\mbox{.}(2023)]%
        {olgun2023dram}
\bibfield{author}{\bibinfo{person}{Ataberk Olgun} {et~al\mbox{.}}} \bibinfo{year}{2023}\natexlab{}.
\newblock \showarticletitle{DRAM Bender: An Extensible and Versatile FPGA-based Infrastructure to Easily Test State-of-the-art DRAM Chips}.
\newblock \bibinfo{journal}{\emph{TCAD}} (\bibinfo{year}{2023}).
\newblock


\bibitem[Park et~al\mbox{.}(2020)]%
        {park2020graphene}
\bibfield{author}{\bibinfo{person}{Yeonhong Park} {et~al\mbox{.}}} \bibinfo{year}{2020}\natexlab{}.
\newblock \showarticletitle{Graphene: Strong yet lightweight row hammer protection}. In \bibinfo{booktitle}{\emph{MICRO}}. IEEE, \bibinfo{pages}{1--13}.
\newblock


\bibitem[Qureshi et~al\mbox{.}(2022)]%
        {qureshi2022hydra}
\bibfield{author}{\bibinfo{person}{Moinuddin Qureshi} {et~al\mbox{.}}} \bibinfo{year}{2022}\natexlab{}.
\newblock \showarticletitle{Hydra: enabling low-overhead mitigation of row-hammer at ultra-low thresholds via hybrid tracking}. In \bibinfo{booktitle}{\emph{ISCA}}.
\newblock


\bibitem[Rakin et~al\mbox{.}(2019)]%
        {rakin2019bit}
\bibfield{author}{\bibinfo{person}{Adnan~Siraj Rakin} {et~al\mbox{.}}} \bibinfo{year}{2019}\natexlab{}.
\newblock \showarticletitle{Bit-flip attack: Crushing neural network with progressive bit search}. In \bibinfo{booktitle}{\emph{ICCV}}. \bibinfo{pages}{1211--1220}.
\newblock


\bibitem[Seaborn and Dullien(2015)]%
        {seaborn2015exploiting}
\bibfield{author}{\bibinfo{person}{Mark Seaborn} {and} \bibinfo{person}{Thomas Dullien}.} \bibinfo{year}{2015}\natexlab{}.
\newblock \showarticletitle{Exploiting the DRAM rowhammer bug to gain kernel privileges}.
\newblock \bibinfo{journal}{\emph{Black Hat}}  \bibinfo{volume}{15} (\bibinfo{year}{2015}), \bibinfo{pages}{71}.
\newblock


\bibitem[Seyedzadeh et~al\mbox{.}(2016)]%
        {seyedzadeh2016counter}
\bibfield{author}{\bibinfo{person}{Seyed~Mohammad Seyedzadeh} {et~al\mbox{.}}} \bibinfo{year}{2016}\natexlab{}.
\newblock \showarticletitle{Counter-based tree structure for row hammering mitigation in DRAM}.
\newblock \bibinfo{journal}{\emph{CAL}}  \bibinfo{volume}{16} (\bibinfo{year}{2016}).
\newblock


\bibitem[Seyedzadeh et~al\mbox{.}(2018)]%
        {seyedzadeh2018mitigating}
\bibfield{author}{\bibinfo{person}{Seyed~Mohammad Seyedzadeh} {et~al\mbox{.}}} \bibinfo{year}{2018}\natexlab{}.
\newblock \showarticletitle{Mitigating wordline crosstalk using adaptive trees of counters}. In \bibinfo{booktitle}{\emph{ISCA}}. IEEE, \bibinfo{pages}{612--623}.
\newblock


\bibitem[Son et~al\mbox{.}(2017)]%
        {son2017making}
\bibfield{author}{\bibinfo{person}{Mungyu Son} {et~al\mbox{.}}} \bibinfo{year}{2017}\natexlab{}.
\newblock \showarticletitle{Making DRAM stronger against row hammering}. In \bibinfo{booktitle}{\emph{DAC}}. \bibinfo{pages}{1--6}.
\newblock


\bibitem[Woo et~al\mbox{.}(2022)]%
        {woo2022scalable}
\bibfield{author}{\bibinfo{person}{Jeonghyun Woo} {et~al\mbox{.}}} \bibinfo{year}{2022}\natexlab{}.
\newblock \showarticletitle{Scalable and Secure Row-Swap: Efficient and Safe Row Hammer Mitigation in Memory Systems}.
\newblock \bibinfo{journal}{\emph{preprint arXiv:2212.12613}} (\bibinfo{year}{2022}).
\newblock


\bibitem[Yao et~al\mbox{.}(2020)]%
        {yao2020deephammer}
\bibfield{author}{\bibinfo{person}{Fan Yao} {et~al\mbox{.}}} \bibinfo{year}{2020}\natexlab{}.
\newblock \showarticletitle{Deephammer: Depleting the intelligence of deep neural networks through targeted chain of bit flips}. In \bibinfo{booktitle}{\emph{USENIX}}.
\newblock


\bibitem[Zhou et~al\mbox{.}(2022a)]%
        {zhou2022lt}
\bibfield{author}{\bibinfo{person}{Ranyang Zhou} {et~al\mbox{.}}} \bibinfo{year}{2022}\natexlab{a}.
\newblock \showarticletitle{LT-PIM: An LUT-Based Processing-in-DRAM Architecture With RowHammer Self-Tracking}.
\newblock \bibinfo{journal}{\emph{IEEE Computer Architecture Letters}} \bibinfo{volume}{21}, \bibinfo{number}{2} (\bibinfo{year}{2022}), \bibinfo{pages}{141--144}.
\newblock


\bibitem[Zhou et~al\mbox{.}(2022b)]%
        {zhou2022red}
\bibfield{author}{\bibinfo{person}{Ranyang Zhou} {et~al\mbox{.}}} \bibinfo{year}{2022}\natexlab{b}.
\newblock \showarticletitle{ReD-LUT: Reconfigurable in-DRAM LUTs enabling massive parallel computation}. In \bibinfo{booktitle}{\emph{Proceedings of the 41st IEEE/ACM International Conference on Computer-Aided Design}}. \bibinfo{pages}{1--8}.
\newblock


\bibitem[Zhou et~al\mbox{.}(2023a)]%
        {zhou2023dnn}
\bibfield{author}{\bibinfo{person}{Ranyang Zhou} {et~al\mbox{.}}} \bibinfo{year}{2023}\natexlab{a}.
\newblock \showarticletitle{DNN-Defender: An in-DRAM Deep Neural Network Defense Mechanism for Adversarial Weight Attack}.
\newblock \bibinfo{journal}{\emph{arXiv preprint arXiv:2305.08034}} (\bibinfo{year}{2023}).
\newblock


\bibitem[Zhou et~al\mbox{.}(2023b)]%
        {zhou2023dram}
\bibfield{author}{\bibinfo{person}{Ranyang Zhou} {et~al\mbox{.}}} \bibinfo{year}{2023}\natexlab{b}.
\newblock \showarticletitle{DRAM-Locker: A General-Purpose DRAM Protection Mechanism against Adversarial DNN Weight Attacks}.
\newblock \bibinfo{journal}{\emph{arXiv preprint arXiv:2312.09027}} (\bibinfo{year}{2023}).
\newblock


\bibitem[Zhou et~al\mbox{.}(2023c)]%
        {zhou2023p}
\bibfield{author}{\bibinfo{person}{Ranyang Zhou} {et~al\mbox{.}}} \bibinfo{year}{2023}\natexlab{c}.
\newblock \showarticletitle{P-pim: A parallel processing-in-dram framework enabling row hammer protection}. In \bibinfo{booktitle}{\emph{DATE}}. IEEE, \bibinfo{pages}{1--6}.
\newblock


\end{thebibliography}
